\documentclass[conference]{IEEEtran}
\IEEEoverridecommandlockouts

\ifCLASSINFOpdf

\else

\fi
\usepackage{siunitx}
\usepackage{cite}
\usepackage{amsmath,amssymb,amsfonts}
\usepackage{algorithmic}
\usepackage{graphicx}
\usepackage{textcomp}
\usepackage{xcolor}
\usepackage{svg}
\usepackage{nomencl}
\usepackage{booktabs}
\usepackage{amsmath,amssymb,amsfonts}
\usepackage{array}
\usepackage{booktabs}
\usepackage{cite}
\usepackage{url}
\usepackage{svg}
\usepackage{graphicx}
\usepackage{tabularx}
\usepackage{booktabs}
\usepackage{siunitx}
\usepackage{threeparttable}

\usepackage{comment}
\def\BibTeX{{\rm B\kern-.05em{\sc i\kern-.025em b}\kern-.08emT\kern-.1667em\lower.7ex\hbox{E}\kern-.125emX}}

\title{Degradation Aware V2G Dispatch for Co-Optimization of Parking Facility and Edge Data Center Located in the Airport Campus}

\author{
\IEEEauthorblockN{Abdullah Al Mehadi\textsuperscript{1},
Hamed Haggi\textsuperscript{2}, \textit{Member, IEEE}, and
Chinmay Morankar\textsuperscript{2}, \textit{Senior Member, IEEE}}
\IEEEauthorblockA{\textsuperscript{1}Dept. of Electrical and Computer Engineering,
Utah State University, Logan, Utah, USA}
\IEEEauthorblockA{\textsuperscript{2}WSP USA Inc., Houston, TX, USA\\
Email: abdullah.mehadi@usu.edu, \{hamed.haggi, chinmay.morankar\}@wsp.com}
}

\begin{document}
\maketitle


\begin{abstract} This paper studies an airport campus where long stay EV parking, airport buildings, and an AI data center share a single grid connection. The campus pays a commercial electricity tariff that charges for both the energy used and the highest power drawn during peak hours. Parked EVs can send power back to the campus through vehicle to grid (V2G) chargers and lower this peak. This paper proposes a two stage method. In the first stage, a linear program schedules EV charging and discharging to maximize the operator's profit and also sets a fixed daily payment to EV owners. In the second stage, a search chooses the share of energy savings paid to owners. The model guarantees that every owner is paid enough to take part, pays for battery wear so that EVs discharge only when it is worth the wear, and assigns charging ports on a first come, first served basis. In the case study, 400 V2G chargers are added to 1200 existing Level 2 chargers, serving 5000 expected sessions under summer and winter rates. In a summer week, V2G raises operator profit by \$4978 compared with smart charging and lowers the high peak demand by 4400~kW (9\,\%). The best contract gives owners 70\,\% of the energy savings and earns 21\,\% more than a fixed payment alone. Doubling the V2G ports to 800 lowers the peak further but reduces profit, because owner payments grow faster than the savings. V2G pays back within 10 years if a bidirectional charger costs less than about \$1800 more than a standard one. Compared with a 5~MW/10~MWh stationary battery, V2G cuts more peak demand at \$409 of investment per kW instead of \$1522 and needs no extra land, while the battery does not pay back within its lifetime. \end{abstract}

\vspace{0.3 cm}


\section*{Nomenclature}
\addcontentsline{toc}{section}{Nomenclature}
\noindent\textit{Sets and Indices}
\begin{IEEEdescription}[\IEEEsetlabelwidth{$\Pi^{\mathrm{noV2G}}_{\mathrm{op}}$\;}]
\setlength{\itemsep}{1pt}
\item[$t$]            Time step, $t\in\{1,\dots,T\}$.
\item[$e$]            Index over all EVs.
\item[$i$]            Index over V2G EVs, $i\in\mathcal{M}$.
\item[$p$]            Index over charger ports.
\item[$\mathcal{M}$]  Set of V2G capable EVs.
\item[$\mathcal{A}$]  Candidate set of energy shares.
\item[$\mathrm{HP},\mathrm{LP}$] High and low peak demand windows.
\item[$p(t)$]         TOU period of step $t$, $\in\{\mathrm{HP},\mathrm{LP},\mathrm{base}\}$.
\end{IEEEdescription}

\vspace{0.2 cm}
\noindent\textit{Decision Variables}
\begin{IEEEdescription}[\IEEEsetlabelwidth{$\Pi^{\mathrm{noV2G}}_{\mathrm{op}}$\;}]
\setlength{\itemsep}{1pt}
\item[$g_t$]                 Grid import at step $t$ (kW).
\item[$p^{\mathrm{ch}}_{e,t}$]  Charging power of EV $e$ (kW).
\item[$p^{\mathrm{dch}}_{i,t}$] Discharging power of V2G EV $i$ (kW).
\item[$\mathrm{soc}_{e,t}$]  Stored energy of EV $e$ (kWh).
\item[$P^{\mathrm{fac}}$]    Facilities peak demand (kW).
\item[$P^{\mathrm{HP}}$] High peak demand (kW).
\item[$P^{\mathrm{LP}}$] Low peak demand (kW).
\item[$F,\alpha$]            Contract: fixed credit (\$/EV-day) and energy share.
\item[$b^{\mathrm{ch}}_t$, $b^{\mathrm{dch}}_t$] BESS charge/discharge power (kW).
\item[$e^{\mathrm{B}}_t$]    BESS stored energy (kWh).
\end{IEEEdescription}

\vspace{0.2 cm}
\noindent\textit{Parameters}
\begin{IEEEdescription}[\IEEEsetlabelwidth{$\Pi^{\mathrm{noV2G}}_{\mathrm{op}}$\;}]
\setlength{\itemsep}{1pt}
\item[$\lambda_t$]           A-3 energy price at step $t$ (\$/kWh).
\item[$\lambda^{\mathrm{TOU}}_{p(t)}$] TOU rate of period $p(t)$ (\$/kWh).
\item[$\lambda^{\mathrm{ECA}}$] Energy cost adder (\$/kWh).
\item[$\Delta t$]            Time step length (h).
\item[$D$]                   Horizon length (days), $D=7$.
\item[$s$]                   Demand charge proration, $s=D/30$.
\item[$\delta_{\mathrm{EV}}$] Commercial EV energy credit (\$/kWh).
\item[$c_{\mathrm{fac}}$]    Facilities demand rate (\$/kW-mo).
\item[$c_{\mathrm{HP}}$] High peak demand rate (\$/kW-mo).
\item[$c_{\mathrm{LP}}$] Low peak demand rate (\$/kW-mo).
\item[$c_{\mathrm{svc}}$]    Service charge (\$/mo).
\item[$c_{\mathrm{deg},i}$]  Wear cost of EV $i$ (\$/kWh discharged).
\item[$r_t$]                 Retail price for non-V2G sessions (\$/kWh).
\item[$d_i$]                 Parked duration of EV $i$ (days).
\item[$L^{\mathrm{dc}}_t$] Data center load (kW).
\item[$L^{\mathrm{bldg}}_t$] Building load (kW).
\item[$\bar{P}^{\mathrm{ch}}_{e}$] Charge power rating of EV $e$ (kW).
\item[$\bar{P}^{\mathrm{dch}}_{i}$] Discharge power rating of EV $i$ (kW).
\item[$A_{e,t}$]             Plugged in indicator, $\in\{0,1\}$.
\item[$M_e$]                 Big-$M$ bound, $\max(\bar{P}^{\mathrm{ch}}_{e},\bar{P}^{\mathrm{dch}}_{e})$.
\item[$\eta_{\mathrm{ch}}$, $\eta_{\mathrm{dch}}$] Charge/discharge efficiencies.
\item[$Q_e$]                 Battery capacity of EV $e$ (kWh).
\item[$\mathrm{cap}_i$]      Usable capacity of V2G EV $i$ (kWh).
\item[$\mathrm{SOC}_{\min}$] Minimum SOC (fraction).
\item[$\mathrm{SOC}_{\max}$] Maximum SOC (fraction).
\item[$a_e,d_e$]             Arrival and departure steps of EV $e$.
\item[$\mathrm{soc}^{0}_{e}$] Arrival SOC of EV $e$ (fraction).
\item[$\mathrm{soc}^{\mathrm{tgt}}_{e}$] Departure target SOC of EV $e$ (fraction).
\item[$N_{\max}$]            Discharge cycle limit over the horizon.
\item[$\mathrm{DoD}$]        Reference depth of discharge.
\item[$\rho$]                Risk premium in the participation reserve.
\item[$W_{\mathrm{s}},W_{\mathrm{w}}$] Summer and winter weeks per year.
\end{IEEEdescription}

\vspace{0.2 cm}
\noindent\textit{Battery Degradation}
\begin{IEEEdescription}[\IEEEsetlabelwidth{$\Pi^{\mathrm{noV2G}}_{\mathrm{op}}$\;}]
\setlength{\itemsep}{1pt}
\item[$\hat V,\hat z$]       Mean cell voltage (V) and mean SOC.
\item[$\beta$]               Cyclic aging factor.
\item[$c_{\mathrm{Ah}}$]     kWh to Ah conversion factor.
\item[$Q^{\mathrm{EOL}}_i$]  End of life charge throughput (Ah).
\item[$c^{\mathrm{B}}_{\mathrm{deg}}$] BESS wear cost (\$/kWh).
\end{IEEEdescription}

\vspace{0.2 cm}
\noindent\textit{Cost Terms and Payoffs}
\begin{IEEEdescription}[\IEEEsetlabelwidth{$\Pi^{\mathrm{noV2G}}_{\mathrm{op}}$\;}]
\setlength{\itemsep}{1pt}
\item[$C_{\mathit{energy}}$] Grid energy cost (\$).
\item[$C_{\mathit{EV}}$]     Commercial EV credit (\$).
\item[$C_{\mathit{demand}}$] Demand charges (\$).
\item[$C_{\mathit{pay}}$]    Payments to V2G owners (\$).
\item[$R_{\mathit{ret}}$]    Dispatch dependent retail revenue (\$).
\item[$B^{\mathrm{A3}}$]     Total A-3 bill over the horizon (\$).
\item[$E^{\mathrm{EV}}$]     EV energy eligible for the EV credit (kWh).
\item[$E^{\mathrm{dch}}_i$]  Energy discharged by EV $i$ (kWh).
\item[$E^{\mathrm{max}}_i$] Largest allowed discharge of EV $i$ (kWh).
\item[$\Pi_{\mathrm{op}}$] Operator profit with V2G (\$).
\item[$\Pi^{\mathrm{noV2G}}_{\mathrm{op}}$] Operator profit without V2G (\$).
\item[$\Delta\Pi$]           Profit gain over smart charging (\$).
\item[$\pi_i$]               Net payoff to owner $i$ (\$).
\item[$U_{\mathrm{res},i}$]  Participation reserve of owner $i$ (\$).
\end{IEEEdescription}

\vspace{0.2 cm}
\noindent\textit{Charger Port Queueing}
\begin{IEEEdescription}[\IEEEsetlabelwidth{$\Pi^{\mathrm{noV2G}}_{\mathrm{op}}$\;}]
\setlength{\itemsep}{1pt}
\item[$N^{\mathrm{V2G}}$]  Number of bidirectional V2G ports.
\item[$N^{\mathrm{L2}}$]   Number of Level~2 ports.
\item[$N^{\mathrm{DCFC}}$] Number of DC fast charge ports.
\item[$H^{\min}_{e}$]      Fewest full power steps EV $e$ needs.
\item[$\phi_p$]            Earliest step at which port $p$ is free.
\item[$a'_{e}$]            Assigned start step of EV $e$.
\item[$w_e$]               Waiting time of EV $e$ (steps).
\item[$w_{\max}$]          Longest allowed wait (steps).
\item[$T$]                 Horizon length (steps).
\item[$b_{p,t}$]           Port busy indicator, $\in\{0,1\}$.
\item[$\mathrm{occ}(t)$]   Number of busy ports at step $t$.
\end{IEEEdescription}

\section{Introduction}

Transportation is moving to electric power, and data centers are growing at the same time. Both add large loads at the same locations on the distribution network. At large commercial sites such as airports, these two loads often share one grid connection. Their combined peak raises the site's demand charges and exposes the operator to time of use (TOU) energy prices. At the same time, electric vehicles (EVs) that stay parked at such sites for days form a large pool of batteries that would otherwise sit idle. Using these batteries through vehicle to grid (V2G) operation can flatten the site peak, shift energy to cheaper hours, and delay the purchase of stationary storage.

Two linked problems must be solved to use this potential. The first is operational: thousands of different EVs must be scheduled together with the data center and building loads under a real commercial tariff \cite{li2026collaborative}, while each vehicle's arrival state of charge, departure target, parking time, and charger availability are respected. The second is economic: EV owners will only let their batteries be discharged if they are paid for the battery wear and for the trouble of taking part \cite{sen2026consent}. A payment scheme that ignores battery wear either cycles the fleet too much, which hurts owners, or fails to attract them at all \cite{calearo2022grid}.

Rahman et al. \cite{rahman2023making} designed fixed term V2G contracts together with online EV scheduling for virtual power plants that take part in electricity markets. Their method raised profit while still meeting each owner's charging need before departure. Orfanoudakis et al. \cite{orfanoudakis2024ev2gym} built EV2Gym, an open source V2G simulator that includes EV, charger, transformer, user behavior, and battery wear models for smart charging and reinforcement learning research. Brinkel et al. \cite{brinkel2020should} built a multi objective V2G model with battery wear and transformer limits that cut costs by 32.4\% and emissions by 23.6\%, and questioned whether grid reinforcement is needed. In \cite{brinkel2026can}, the same group compared six grid tariff designs for EV charging and found that capacity based tariffs lower grid expansion costs and are fairer, but that no single design is best for efficiency, fairness, and predictability at once.

Only a few studies look at EV scheduling in airport parking. The ALIGHT report \cite{thingvad2023potential} studied smart charging and V2G at Copenhagen Airport. It found frequency reserve (FCR-D) to be the most profitable service, worth about €250--€350 per year per destination charger and about €2,500 per well used fast charger, with little effect on users. A study at Harry Reid International Airport compared MILP and deep reinforcement learning (DDPG) for EV scheduling with V2G \cite{al2024comparative}. MILP gave higher profit and customer satisfaction, while reinforcement learning was faster and caused less battery wear but handled the peak less well. Guo et al. \cite{guo2023infrastructure} planned an airport microgrid with electric aircraft, parking lot EVs, and V2G using NSGA-II, and showed that V2G lowers operating cost. Vereno et al. \cite{vereno2025utilizing} showed that automated airport garages can provide fast frequency reserve while still fully charging cars, but that regulatory approval is the main barrier. Yan et al. \cite{yan2026optimal} combined LSTM forecasting and MILP to size PV, ground service vehicle fleets, and chargers at Guangzhou Baiyun airport, cutting annual cost by 624,042.60~CNY with a 3.04 year payback.

None of these studies combines an airport parking fleet with a colocated AI data center under a demand charge tariff, or designs the owner payment as part of the dispatch. The main contributions of this paper are:

\begin{itemize}
\item An airport campus is modeled in which long stay EV parking, airport buildings, and a real trace AI data center share one grid connection under a commercial tariff with both energy and demand charges. Earlier airport V2G studies consider the parking load on its own.

\item A two stage method is proposed that designs a two part V2G contract (a fixed daily credit and a share of the energy value) together with the dispatch. Owner participation and battery wear are built into the first stage, so every owner is paid enough to take part and EVs discharge only when the benefit is larger than the wear. Because the fixed credit is found inside the LP, the second stage is a simple one variable search.

\item Each vehicle is modeled with its own arrival state of charge, departure target, and parking time, and is assigned a charger port on a first come, first served basis. A comparison of 400 and 800 bidirectional ports shows that the profitable number of V2G vehicles is set by the demand charges and owner payments, not by the number of ports.

\item V2G is compared with a stationary BESS as an investment, using the same optimization, both tariff seasons, and the land cost of the BESS.
\end{itemize}

\begin{figure}[t]
\centering
\includegraphics[width=\columnwidth]{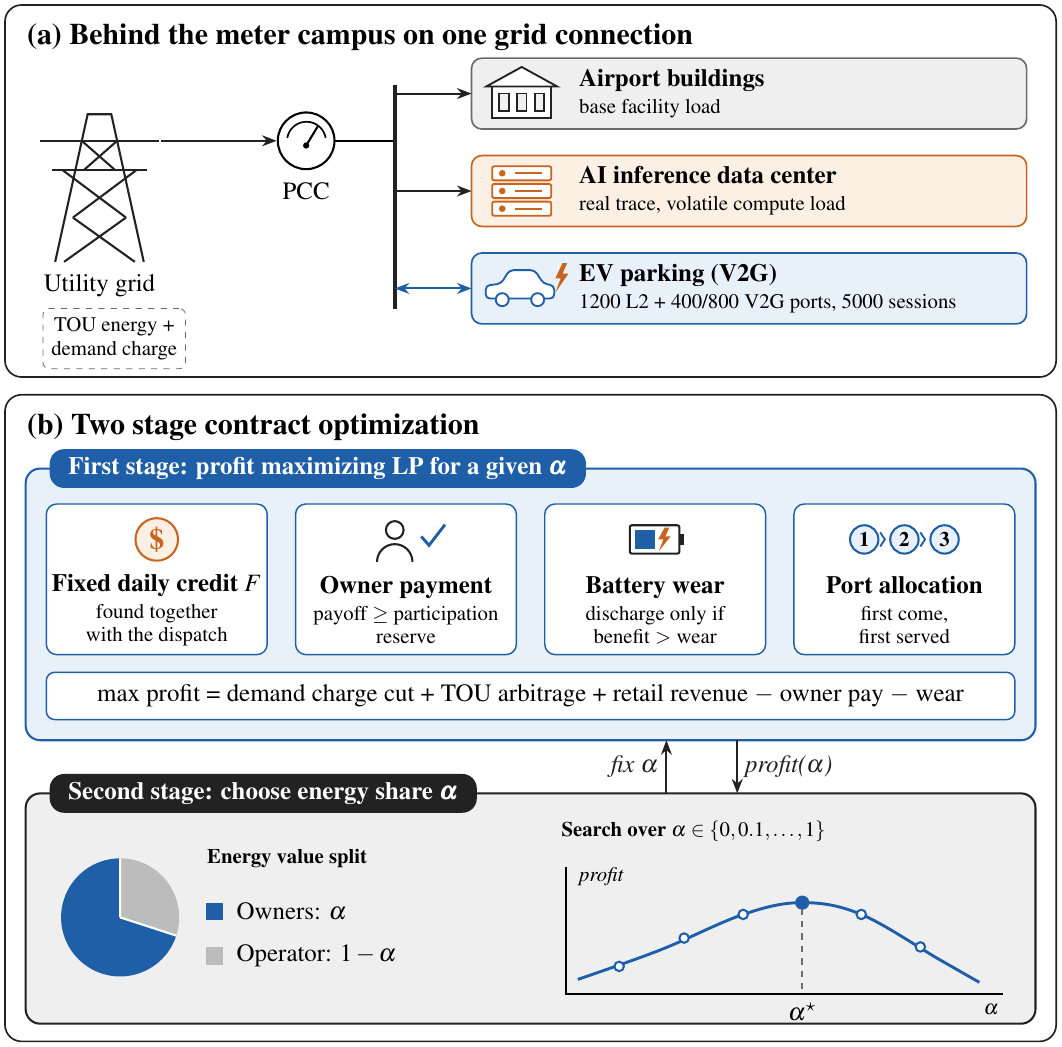}
\caption{Overview of the campus and the two stage method.}
\label{fig:clipart}
\end{figure}

The rest of the paper is organized as follows. Section~II describes the method, Section~III presents the case study and results, and Section~IV concludes the paper. The overall method is shown in Fig.~\ref{fig:clipart}.

\section{Methodology}

\subsection{Two Stage Framework}
\label{subsec:framework}
The proposed framework has two stages, as shown in Fig.~\ref{fig:clipart}. The first stage is a linear program (LP) that schedules all EVs for the week ahead, together with the data center and building loads, so that the operator's net profit is maximized. For a given energy share $\alpha$, this stage also chooses the fixed daily credit $F$ paid to V2G owners. The second stage searches over $\alpha$ and keeps the value that gives the highest profit. Together, $(F,\alpha)$ form the two part V2G contract offered to owners.

\subsection{First Stage: Dispatch and Fixed Credit}
\label{subsec:stage1}
The first stage minimizes the operator's net cost, which is the A-3 bill of \eqref{eq:a3_bill} plus the owner payments, minus the retail revenue that depends on the dispatch,
\begin{equation}
\min \; C_{\mathit{energy}} - C_{\mathit{EV}} + C_{\mathit{demand}} + C_{\mathit{pay}} - R_{\mathit{ret}},
\label{eq:obj_total}
\end{equation}
where each term is
\begin{align}
C_{\mathit{energy}} &= \sum_{t} \lambda_t\, g_t\, \Delta t, \label{eq:c_energy}\\
C_{\mathit{EV}}     &= \delta_{\mathrm{EV}} \sum_{e}\sum_{t} p^{\mathrm{ch}}_{e,t}\, \Delta t, \label{eq:c_ev}\\
C_{\mathit{demand}} &= s\!\left(c_{\mathrm{fac}}P^{\mathrm{fac}} + c_{\mathrm{HP}}P^{\mathrm{HP}} + c_{\mathrm{LP}}P^{\mathrm{LP}}\right), \label{eq:c_demand}\\
C_{\mathit{pay}}    &= \sum_{i\in\mathcal{M}}\!\Big[F d_i + \sum_{t}\left(\alpha\lambda_t + c_{\mathrm{deg},i}\right) p^{\mathrm{dch}}_{i,t}\,\Delta t\Big], \label{eq:c_pay}\\
R_{\mathit{ret}}    &= \sum_{e\notin\mathcal{M}}\sum_{t} r_t\, p^{\mathrm{ch}}_{e,t}\,\Delta t. \label{eq:r_ret}
\end{align}

The owner payment $C_{\mathit{pay}}$ has three parts: a fixed credit for each parked day, a share $\alpha$ of the energy value of each discharged kWh, and a wear payment for each discharged kWh. The wear coefficient $c_{\mathrm{deg},i}$ is derived in Section~\ref{subsec:deg}. Non-V2G sessions pay the retail price $r_t$, which is \$0.54/kWh in the A-3 high peak window and \$0.45/kWh otherwise. V2G owners pay a flat \$0.45/kWh only for the energy they actually need, $(\mathrm{soc}^{\mathrm{tgt}}_{i}-\mathrm{soc}^{0}_{i})Q_i/\eta_{\mathrm{ch}}$. This is the same amount they would pay without V2G, so an owner is never billed for energy that the operator discharges and later recharges. This revenue and the \$2 per session fee do not depend on the dispatch, so they are left out of \eqref{eq:obj_total} but are included in the reported profit. As a result, the first stage, the second stage, and the results in Section~\ref{sec:results} all use the same profit measure.

The grid import must cover both site loads and the net battery charging. Export to the grid is not allowed,
\begin{equation}
g_t = L^{\mathrm{dc}}_t + L^{\mathrm{bldg}}_t + \sum_{e}\!\left(p^{\mathrm{ch}}_{e,t} - p^{\mathrm{dch}}_{e,t}\right) + b^{\mathrm{ch}}_t - b^{\mathrm{dch}}_t,
\quad g_t \ge 0.
\label{eq:balance}
\end{equation}
The BESS terms $b^{\mathrm{ch}}_t,b^{\mathrm{dch}}_t$ are zero except in the BESS benchmark of Section~\ref{subsec:bess}. The peak variables track the highest import in each demand window,
\begin{equation}
P^{\mathrm{fac}} \ge g_t\;\;\forall t,
\quad P^{\mathrm{HP}} \ge g_t\;\;\forall t\in\mathrm{HP},
\quad P^{\mathrm{LP}} \ge g_t\;\;\forall t\in\mathrm{LP}.
\label{eq:peak}
\end{equation}

Each EV can charge or discharge only while it is plugged in, and only V2G EVs can discharge \cite{al2025milp, al2026benchmarking},
\begin{align}
0 &\le p^{\mathrm{ch}}_{e,t} \le \bar{P}^{\mathrm{ch}}_{e} A_{e,t}, \label{eq:ch_lim}\\
0 &\le p^{\mathrm{dch}}_{i,t} \le \bar{P}^{\mathrm{dch}}_{i} A_{i,t}, \quad i\in\mathcal{M}. \label{eq:dch_lim}
\end{align}
A vehicle cannot charge and discharge at full power in the same step,
\begin{equation}
p^{\mathrm{ch}}_{e,t} + p^{\mathrm{dch}}_{e,t} \le M_e A_{e,t}.
\label{eq:excl}
\end{equation}
The stored energy changes with the charge and discharge efficiencies,
\begin{equation}
\mathrm{soc}_{e,t+1} = \mathrm{soc}_{e,t} + \!\left(\eta_{\mathrm{ch}}\, p^{\mathrm{ch}}_{e,t} - p^{\mathrm{dch}}_{e,t}/\eta_{\mathrm{dch}}\right)\!\Delta t,
\label{eq:soc_dyn}
\end{equation}
and stays within the allowed range,
\begin{equation}
\mathrm{SOC}_{\min} Q_e \le \mathrm{soc}_{e,t} \le \mathrm{SOC}_{\max} Q_e.
\label{eq:soc_bnd}
\end{equation}
Each session starts at its arrival level and must end exactly at its departure target,
\begin{equation}
\mathrm{soc}_{e,a_e} = \mathrm{soc}^{0}_{e} Q_e,
\qquad \mathrm{soc}_{e,d_e+1} = \mathrm{soc}^{\mathrm{tgt}}_{e} Q_e.
\label{eq:session}
\end{equation}
The target is an equality so that the operator cannot raise its retail revenue by charging a vehicle beyond what the owner asked for. The total discharge of each V2G EV is limited by the allowed number of cycles,
\begin{equation}
\sum_{t} p^{\mathrm{dch}}_{i,t}\,\Delta t \le N_{\max}\,\mathrm{DoD}\,\mathrm{cap}_i,
\qquad i\in\mathcal{M}.
\label{eq:cycle}
\end{equation}
Finally, every V2G owner must receive at least a minimum payoff, called the participation reserve,
\begin{equation}
F d_i + \alpha \sum_{t}\lambda_t\, p^{\mathrm{dch}}_{i,t}\,\Delta t \;\ge\; U_{\mathrm{res},i}\;\;\forall i\in\mathcal{M}.
\label{eq:participation}
\end{equation}
The left side is the owner's net payoff, explained in \eqref{eq:owner_payoff}. Because \eqref{eq:participation} is a constraint of the LP, and not a check made after solving, every solution pays every owner enough to take part.

The owner's net payoff is the fixed credit, plus the energy share, plus the wear payment, minus the wear the owner actually suffers,
\begin{equation}
\pi_i = F\, d_i + \alpha \sum_{t} \lambda_t\, p^{\mathrm{dch}}_{i,t}\,\Delta t + c_{\mathrm{deg},i} E^{\mathrm{dch}}_i - c_{\mathrm{deg},i} E^{\mathrm{dch}}_i.
\label{eq:owner_payoff}
\end{equation}
The wear payment cancels the wear cost, so $\pi_i$ equals the left side of \eqref{eq:participation}. The participation reserve is set from the largest wear the owner could face, plus a risk premium $\rho=0.30$ \cite{wang2025electric},
\begin{equation}
\begin{aligned}
U_{\mathrm{res},i} &= (1+\rho)\, c_{\mathrm{deg},i}\, E^{\mathrm{max}}_i,\\
E^{\mathrm{max}}_i &= N_{\max}\,\mathrm{DoD}\,\mathrm{cap}_i.
\end{aligned}
\label{eq:u_res}
\end{equation}

\subsection{Second Stage: Energy Share Search}
\label{subsec:upper_grid}
When $\alpha$ is fixed, \eqref{eq:participation} is linear in $F$ and in the discharge powers. For this reason, $F$ is a decision variable of the first stage LP and is found together with the dispatch. The second stage then only needs to search over one variable, $\alpha$, and picks the value that gives the largest profit gain over smart charging,
\begin{equation}
\alpha^{\star} = \arg\max_{\alpha\in\mathcal{A}} \left[\Pi_{\mathrm{op}}\!\left(F^{\star}(\alpha),\alpha\right) - \Pi^{\mathrm{noV2G}}_{\mathrm{op}}\right].
\label{eq:upper}
\end{equation}
With $\mathcal{A}=\{0,0.1,\dots,1\}$, the contract is found from 11 LP solves instead of a full search over a two dimensional $(F,\alpha)$ grid. Since every solve enforces \eqref{eq:participation}, all candidate contracts are acceptable to the owners.

\subsection{LADWP A-3 Commercial Tariff}
\label{subsec:ad3}
The campus meter is billed under the LADWP A-3(A) time of use (TOU) schedule \cite{LADWP_EV_Rates}. The bill has an energy part, a demand part, and a fixed service part. The energy price in each step is the seasonal TOU rate plus a fixed energy cost adjustment (ECA),
\begin{equation}
\lambda_t = \lambda^{\mathrm{TOU}}_{p(t)} + \lambda^{\mathrm{ECA}},
\label{eq:a3_total}
\end{equation}
\begin{equation}
\lambda^{\mathrm{TOU}}_{p(t)} =
\begin{cases}
\lambda_{\mathrm{HP}}, & \text{weekday, } 13{:}00\!\le\! t \!<\! 17{:}00,\\
\lambda_{\mathrm{LP}}, & \text{weekday, } t\in[10,13)\cup[17,20),\\
\lambda_{\mathrm{base}}, & \text{otherwise.}
\end{cases}
\label{eq:a3_tou}
\end{equation}
The total A-3 bill over the horizon is
\begin{equation}
\begin{aligned}
B^{\mathrm{A3}} =\;& \sum_{t} \lambda_t\, g_t\, \Delta t
 + s\!\left(c_{\mathrm{fac}} P^{\mathrm{fac}} + c_{\mathrm{HP}} P^{\mathrm{HP}} + c_{\mathrm{LP}} P^{\mathrm{LP}}\right) \\
 &+ s\, c_{\mathrm{svc}} - \delta_{\mathrm{EV}}\, E^{\mathrm{EV}}.
\end{aligned}
\label{eq:a3_bill}
\end{equation}
The rates for both seasons are listed in Section~\ref{subsec:setup}. Annual values are found by weighting the weekly results by the number of weeks in each season,
\begin{equation}
\Delta\Pi^{\mathrm{yr}} = W_{\mathrm{s}}\,\Delta\Pi^{\mathrm{wk}}_{\mathrm{s}} + W_{\mathrm{w}}\,\Delta\Pi^{\mathrm{wk}}_{\mathrm{w}},
\label{eq:annual}
\end{equation}
with $W_{\mathrm{s}}=122/7\approx17.4$ and $W_{\mathrm{w}}=365/7-W_{\mathrm{s}}\approx34.7$.

\subsection{Battery Degradation Model}
\label{subsec:deg}
The wear cost follows the linearized Ecker--Schmalstieg aging model \cite{orfanoudakis2024ev2gym, lee2023vehicle}, with $108\times95$ cells per pack. The mean cell voltage is taken as a linear function of the mean state of charge,
\begin{equation}
\hat V = V_{\min}+\kappa\,\hat z ,
\label{eq:vhat}
\end{equation}
and the cyclic aging factor is
\begin{equation}
\beta = \zeta_0(\hat V-\zeta_1)^2+\zeta_2+\zeta_3\,\Delta\mathrm{DoD} .
\label{eq:beta}
\end{equation}
The optimization works in kWh while the aging model works in Ah, so the pack level conversion factor is
\begin{equation}
c_{\mathrm{Ah}} = \frac{1000}{N_{\mathrm{s}}\,V_{\mathrm{cell}}} \approx 2.57\,\tfrac{\mathrm{Ah}}{\mathrm{kWh}} .
\label{eq:cah}
\end{equation}
The total charge a pack can pass before the end of its life is
\begin{equation}
Q^{\mathrm{EOL}}_i = N_{\mathrm{cyc}}\,\Delta\mathrm{DoD}\, c_{\mathrm{Ah}}\, \mathrm{cap}_i ,
\label{eq:qacc}
\end{equation}
and the wear cost per kWh discharged is
\begin{equation}
c_{\mathrm{deg},i} = 0.5\,\beta\,
\frac{N_{\max}\,c_{\mathrm{Ah}}\,c_{\mathrm{rep}}\,\mathrm{cap}_i}{\sqrt{Q^{\mathrm{EOL}}_i}} .
\label{eq:degcoef}
\end{equation}
This coefficient enters the objective through the wear payment in \eqref{eq:c_pay}. The payment is linear in the discharged energy, so the first stage stays an LP. Because wear is priced inside the objective, the cycle limit is set to its largest value, $N_{\max}=3$, instead of a fixed number of cycles as is common in the literature. An EV then discharges only when the campus benefit is larger than the wear cost.

\subsection{Stationary BESS Benchmark}
\label{subsec:bess}
To test whether the parked fleet can replace owned storage, a benchmark case is solved with no V2G and with a stationary BESS behind the same meter, connected through \eqref{eq:balance}. Its operation is limited by
\begin{equation}
\begin{aligned}
&b^{\mathrm{ch}}_t + b^{\mathrm{dch}}_t \le \bar{P}^{\mathrm{B}}, \qquad b^{\mathrm{ch}}_t,\,b^{\mathrm{dch}}_t \ge 0,\\
&e^{\mathrm{B}}_{t+1} = e^{\mathrm{B}}_{t} + \left(\eta_{\mathrm{B}}\, b^{\mathrm{ch}}_t - b^{\mathrm{dch}}_t/\eta_{\mathrm{B}}\right)\Delta t,\\
&\underline{s}_{\mathrm{B}}\bar{E}^{\mathrm{B}} \le e^{\mathrm{B}}_{t} \le \overline{s}_{\mathrm{B}}\bar{E}^{\mathrm{B}}, \qquad e^{\mathrm{B}}_{0} = e^{\mathrm{B}}_{T} = s^{0}_{\mathrm{B}}\bar{E}^{\mathrm{B}}.
\end{aligned}
\label{eq:bess}
\end{equation}
The end condition stops the BESS from discharging energy that it did not store during the week. BESS wear is added to \eqref{eq:obj_total} as $c_{\mathrm{deg}}^{\mathrm{B}}\sum_t b^{\mathrm{dch}}_t\Delta t$, with $c_{\mathrm{deg}}^{\mathrm{B}} = c^{\mathrm{B}}_{E}/\big(N^{\mathrm{B}}_{\mathrm{cyc}}(\overline{s}_{\mathrm{B}}-\underline{s}_{\mathrm{B}})\big)$. The BESS is solved for both seasons and compared with V2G using the same profit measure.

\subsection{Charger Port Queueing}
\label{subsec:queue}
Charging demand is served from three separate port pools: $N^{\mathrm{V2G}}$ bidirectional ports for V2G vehicles, $N^{\mathrm{L2}}$ Level~2 ports for commuters, and $N^{\mathrm{DCFC}}$ fast charge ports for fast sessions. In each pool, EVs are served in order of arrival and take the first free port. The smallest number of full power steps that an EV needs to reach its target is
\begin{equation}
H^{\min}_{e} = \left\lceil \frac{\left(\mathrm{soc}^{\mathrm{tgt}}_{e} - \mathrm{soc}^{0}_{e}\right) Q_e}{\eta_{\mathrm{ch}}\, \bar{P}^{\mathrm{ch}}_{e}\, \Delta t} \right\rceil .
\label{eq:hmin}
\end{equation}
The EV starts on the first free port $p$, and its start step and waiting time are
\begin{equation}
a'_{e} = \max\!\left(a_e,\, \phi_p\right),
\qquad w_e = a'_{e} - a_e.
\label{eq:wait}
\end{equation}
An EV is turned away if it waits too long, cannot start within the horizon, or cannot finish charging before it leaves,
\begin{equation}
\text{reject if } \; w_e > w_{\max} \;\; \text{or} \;\; a'_{e} \ge T \;\; \text{or} \;\; (d_e - a'_{e}) < H^{\min}_{e}.
\label{eq:reject}
\end{equation}
The number of busy ports in a pool at each step is
\begin{equation}
\mathrm{occ}(t) = \sum_{p} b_{p,t}.
\label{eq:occ}
\end{equation}

\subsection{Vehicle Availability}
\label{subsec:avail}
An EV can be used only between its arrival and departure,
\begin{equation}
A_{e,t} =
\begin{cases}
1, & a_e \le t < d_e \;\;(\text{vehicle present}),\\
0, & \text{otherwise.}
\end{cases}
\label{eq:avail}
\end{equation}
The arrival and departure profiles and the fleet data are given in Section~\ref{subsec:setup}.

\section{Simulation Results}
\subsection{Case Study Setup and Input Data}
\label{subsec:setup}
The campus has 1200 existing Level~2 chargers and 16 DC fast chargers. In the main case, 400 V2G chargers are added, giving 1600 Level~2 class chargers in total. The study covers 5000 expected EVs over the week of Monday 10 to Sunday 16 July 2023, with first come, first served port allocation. This week is first solved under the A-3 summer rates and then solved again under the winter rates, with its own contract search, so that the seasonal difference comes from the tariff alone. The summer rates apply from June to September and the winter rates from October to May. All tariff, battery, and fleet inputs are listed in Tables~\ref{tab:a3_rates}--\ref{tab:fleet}.

Three operating modes are compared: uncontrolled (greedy) charging, smart charging without V2G, and V2G under the optimized two part contract. A sensitivity analysis then tests the main assumptions, and V2G is finally compared with a stationary BESS as an investment.

\begin{table}[t]
\centering
\caption{LADWP A-3 Tariff Parameters}
\label{tab:a3_rates}
\footnotesize
\renewcommand{\arraystretch}{1.1}
\begin{tabular}{@{}llcc@{}}
\toprule
\textbf{Symbol} & \textbf{Description} & \textbf{Summer} & \textbf{Winter} \\
\midrule
$\lambda_{\mathrm{HP}}$   & High peak energy (\$/kWh)      & 0.05991 & 0.05464 \\
$\lambda_{\mathrm{LP}}$   & Low peak energy (\$/kWh)       & 0.05365 & 0.05464 \\
$\lambda_{\mathrm{base}}$ & Base energy (\$/kWh)           & 0.03356 & 0.03798 \\
$\lambda^{\mathrm{ECA}}$  & Energy cost adder (\$/kWh)     & 0.05690 & 0.05690 \\
$c_{\mathrm{HP}}$         & High peak demand (\$/kW-mo)    & 9.70    & 4.30 \\
$c_{\mathrm{LP}}$         & Low peak demand (\$/kW-mo)     & 3.30    & 0.00 \\
$c_{\mathrm{fac}}$        & Facilities demand (\$/kW-mo)   & 4.56    & 4.56 \\
$c_{\mathrm{svc}}$        & Service charge (\$/mo)         & 75.00   & 75.00 \\
$\delta_{\mathrm{EV}}$    & EV charging credit (\$/kWh)    & 0.025   & 0.025 \\
\bottomrule
\end{tabular}
\end{table}

\begin{table}[t]
\centering
\caption{Degradation Model Parameters}
\label{tab:deg}
\footnotesize
\begin{tabular}{@{}llc@{}}
\toprule
\textbf{Symbol} & \textbf{Description} & \textbf{Value} \\
\midrule
$V_{\min}$ & Cell voltage at $0\%$ SOC & $3.3324$~V \\
$\kappa$ & OCV slope, $\mathrm{OCV}(1)-\mathrm{OCV}(0)$ & $0.8263$~V \\
$\hat z$ & Mean SOC operating point & $0.5$ \\
$\zeta_0,\ \zeta_1$ & Cyclic coefficients & $7.348\times10^{-3}$, $3.667$ \\
$\zeta_2,\ \zeta_3$ & Cyclic coefficients & $7.600\times10^{-4}$, $4.081\times10^{-3}$ \\
$\Delta\mathrm{DoD}$ & Assumed depth of discharge & $0.75$ \\
$N_{\mathrm{s}}$ & Cells in series & $108$ \\
$V_{\mathrm{cell}}$ & Nominal cell voltage & $3.6$~V \\
$N_{\mathrm{cyc}}$ & Rated cycle life & $4000$ \cite{Jason2025BatteryChemistry}\\
$c_{\mathrm{rep}}$ & Pack replacement cost & $98~\$/\mathrm{kWh}$ \cite{bnef2023battery}\\
$N_{\max}$ & Maximum discharge cycle limit & $3$ \\
\bottomrule
\end{tabular}
\end{table}

\begin{table}[t]
\centering
\caption{Stationary BESS Benchmark Parameters}
\label{tab:bess}
\footnotesize
\renewcommand{\arraystretch}{1.2}
\begin{tabular}{@{}llc@{}}
\toprule
\textbf{Symbol} & \textbf{Description} & \textbf{Value} \\
\midrule
$\bar{P}^{\mathrm{B}},\,\bar{E}^{\mathrm{B}}$ & Power / energy rating & 5~MW / 10~MWh \\
$\eta_{\mathrm{B}}$ & One way efficiency & 0.95 \\
$\underline{s}_{\mathrm{B}},\,\overline{s}_{\mathrm{B}},\,s^{0}_{\mathrm{B}}$ & SOC limits / initial SOC & 0.05 / 0.95 / 0.50 \\
$c^{\mathrm{B}}_{P}$ & Power related capital cost & 432.64~\$/kW \\
$c^{\mathrm{B}}_{E}$ & Energy related capital cost & 254.67~\$/kWh \\
$N^{\mathrm{B}}_{\mathrm{cyc}}$ & Rated cycle life & 6000 \\
$c^{\mathrm{B}}_{\mathrm{O\&M}}$ & Fixed O\&M & 4.322~\$/kWh-yr \\
\bottomrule
\end{tabular}
\end{table}

\begin{table}[t]
\centering
\caption{EV Fleet Segments, Sizing, and Availability}
\label{tab:fleet}
\footnotesize
\renewcommand{\arraystretch}{1.2}
\setlength{\tabcolsep}{3pt}
\begin{tabular}{@{}lccccc@{}}
\toprule
\textbf{Segment} & \textbf{Share} & \textbf{Batt. (kWh)} & \textbf{Charger} & \textbf{Arr. SOC} & \textbf{Dwell} \\
\midrule
V2G      & $23\%$ & 60--80 & 11\,kW L2  & 0.30--0.60 & $\sim$4\,d \\
Commuter & $75\%$ & 55--70 & 7.4\,kW L2 & 0.45--0.55 & 8/20\,h \\
DCFC     & $2\%$  & 60--90 & 150\,kW DC & 0.30--0.35 & 1\,h \\
\bottomrule
\end{tabular}
\end{table}

\begin{figure}[t]
  \centering
  \includegraphics[width=0.85\columnwidth]{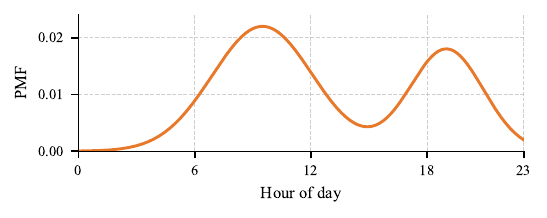}
  \caption{Probability mass function of the EV arrival time.}
  \label{fig:pmf}
\end{figure}

\textit{EV fleet:} The fleet has three groups, listed in Table~\ref{tab:fleet}. Commuters, mostly airport employees, make up the largest group and stay 8 or 20 hours. Long stay V2G vehicles belong to air passengers and stay about four days. Fast charging vehicles belong to people dropping off friends or family and stay about one hour. Arrival and departure times are drawn from a two peak distribution that follows airport activity, with a morning peak near 09:00--10:00 and an evening peak near 21:00 \cite{SFO2023EVLoadAnalysis}, as shown in Fig.~\ref{fig:pmf}.

\textit{Data center load:} The AI data center load is taken from the model of Fan et al. \cite{fan2026co}, which describes a 430 GPU data center. It is scaled up 20 times to represent 8600 GPUs.

\textit{BESS size:} The BESS power rating of 5~MW is chosen to be close to the 4.4~MW that the 400 V2G vehicles can discharge together (400 $\times$ 11~kW). Both resources can then lower the peak by a similar amount, so the comparison is about cost rather than size. The 2 hour duration (10~MWh) is one of the standard durations in the NREL Annual Technology Baseline \cite{nrel2024atb}. The cost of a BESS grows almost in proportion to its size, while the peak reduction shows diminishing returns, as seen for V2G in Section~\ref{sec:comparison}. A larger BESS is therefore not expected to lower its cost per kW of peak reduction.

\subsection{Selected Deployment: 400 V2G Ports}
\label{sec:results}

\begin{table}[t]
\centering
\caption{Operator Cost and Net Cost After Charging Revenue Across Operating Regimes (400 V2G Ports, Summer Week)}
\label{tab:cost_breakdown}
\footnotesize
\setlength{\tabcolsep}{4pt}
\renewcommand{\arraystretch}{1.15}
\begin{threeparttable}
\sisetup{group-separator={,},group-minimum-digits=4,detect-weight=true}
\begin{tabular}{l S[table-format=-6.0] S[table-format=-6.0] S[table-format=-6.0]}
\toprule
\textbf{Component (\$)} & {\textbf{Uncontrolled}} & {\textbf{Smart}} & {\textbf{V2G}} \\
\midrule
TOU energy         & 687545 & 687493 & 686321 \\
Facility demand    & 52556  & 51689  & 49446 \\
TOU demand         & 149109 & 146825 & 135778 \\
Service charge     & 18     & 18     & 18 \\
EV discount credit & -3277  & -3277  & -5154 \\
\midrule
A-3 grid bill      & 885950 & 882748 & 866409 \\
Owner payments     & {--}   & {--}   & 10192 \\
\midrule
\textbf{Operator cost} & \bfseries 885950 & \bfseries 882748 & \bfseries 876601 \\
Station revenue    & 68019  & 70475  & 69306 \\
\textbf{Net cost}  & \bfseries 817931 & \bfseries 812273 & \bfseries 807295 \\
\midrule
Campus peak (kW)   & {49394} & {48580} & {46472} \\
High peak (kW)     & {49076} & {48580} & {44180} \\
\midrule
\multicolumn{4}{@{}l}{\textit{Change vs. uncontrolled (\%)}} \\
Demand charges     & {--} & {$-1.56$} & {$-8.15$} \\
A-3 grid bill      & {--} & {$-0.36$} & {$-2.21$} \\
Net cost           & {--} & {$-0.69$} & {$-1.30$} \\
Campus peak        & {--} & {$-1.65$} & {$-5.92$} \\
High peak          & {--} & {$-1.01$} & {$-9.98$} \\
\midrule
Revenue / operator cost (\%) &{7.68} & {7.98} & {7.91} \\
\bottomrule
\end{tabular}
\begin{tablenotes}[flush]\footnotesize
\item Net cost is the operator cost minus the charging station revenue. The EV discount credit is a negative entry. Owner payments include the \$4508 wear payment and are part of the V2G operator cost. Demand charges are the facility plus TOU demand rows.
\end{tablenotes}
\end{threeparttable}
\end{table}

\begin{figure}[t]
\centering
\includegraphics[width=\columnwidth]{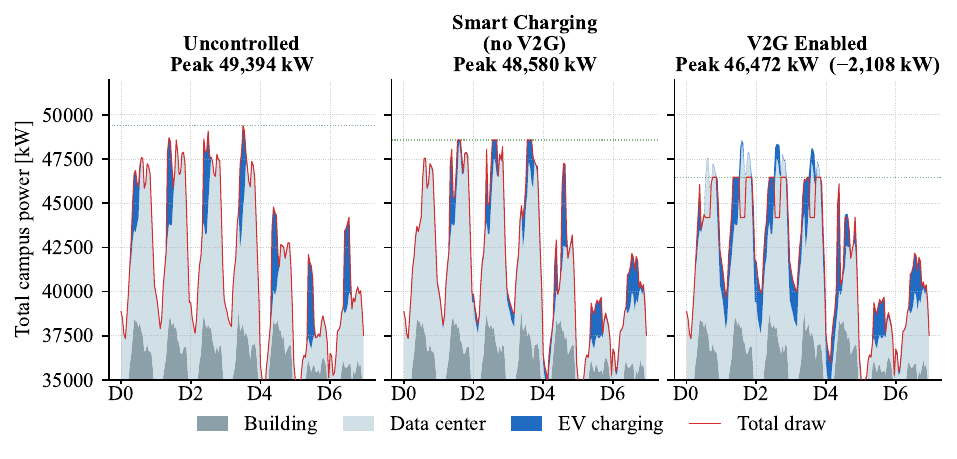}
\caption{Total campus power over the 7 day horizon under uncontrolled, smart charging, and V2G enabled operation (400 ports).}
\label{fig:campus_power}
\end{figure}

\begin{figure}[t]
\centering
\includegraphics[width=\columnwidth]{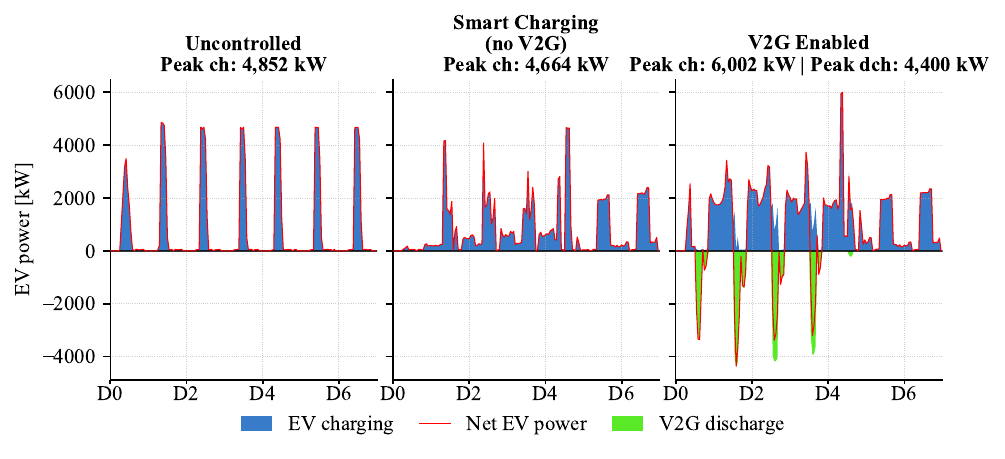}
\caption{EV charging/discharging power profiles for the three operating regimes (400 ports).}
\label{fig:ev_power}
\end{figure}

Table~\ref{tab:cost_breakdown} gives the operator cost, the charging station revenue, and the net cost for the three operating modes over the summer week. Figs.~\ref{fig:campus_power} and \ref{fig:ev_power} show the campus and EV power profiles. In total, 4250 sessions are served, of which 400 are V2G vehicles. Non-V2G sessions pay \$0.54/kWh in the weekday high peak window (1:00--5:00~PM), which is 20\,\% more than the \$0.45/kWh off peak price. V2G vehicles pay a flat \$0.45/kWh only for the energy they need. The net cost, which is the operator cost minus the station revenue, is used as the bottom line.

Smart charging lowers the net cost by \$5658 (0.69\,\%) compared with uncontrolled charging. The grid bill falls by \$3202 (0.36\,\%), mostly because of lower demand charges: the campus peak drops from 49394~kW to 48580~kW, and the facility and TOU demand charges fall by \$866 and \$2284. The other \$2456 comes from higher station revenue (+3.61\,\%). Because the dispatch maximizes profit, it places some non-V2G charging in the high peak window to earn the 20\,\% price premium. This raises the high peak energy bill, but the extra revenue is larger, and the high peak demand still stays at the campus peak of 48580~kW.

V2G lowers the net cost by a further \$4978 (0.61\,\%) compared with smart charging, and by \$10636 (1.30\,\%) compared with uncontrolled charging. The grid bill falls by \$16339 (1.85\,\%) against smart charging. Of this saving, \$10192 (62\,\%) is paid to the vehicle owners, including \$4508 of wear payment, and \$1169 of station revenue is lost because less non-V2G charging falls in the premium window. Again, most of the saving comes from demand charges. The TOU and facility demand charges fall by \$11047 and \$2243 (6.7\,\% in total) against smart charging, because the high peak demand drops by 4400~kW (9.1\,\%), from 48580~kW to 44180~kW. The campus peak falls to 46472~kW, which is 2923~kW below the uncontrolled case. Fig.~\ref{fig:campus_power} shows V2G shaving the daily peaks, and Fig.~\ref{fig:ev_power} shows the daily charge and discharge cycles, with 67761~kWh discharged by the V2G fleet over the week.

\begin{table}[t]
\centering
\caption{Energy Share Search Under the Summer Tariff (400 Ports)}
\label{tab:alpha}
\footnotesize
\setlength{\tabcolsep}{6pt}
\renewcommand{\arraystretch}{1.1}
\begin{tabular}{@{}cccc@{}}
\toprule
$\alpha$ & $F^{\star}(\alpha)$ (\$/EV-day) & $\Delta\Pi$ vs. smart (\$/wk) & Binding owners \\
\midrule
0.0 & 3.70 & 4111 & 19 \\
0.2 & 2.67 & 4359 & 19 \\
0.4 & 1.65 & 4614 & 25 \\
0.6 & 0.63 & 4863 & 63 \\
\textbf{0.7} & \textbf{0.12} & \textbf{4978} & \textbf{208} \\
0.8 & 0.00 & 4464 & 208 \\
1.0 & 0.00 & 3214 & 400 \\
\bottomrule
\end{tabular}
\end{table}

\begin{figure}[t]
\centering
\includegraphics[width=\columnwidth]{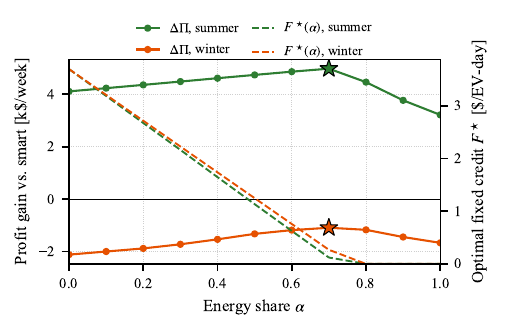}
\caption{Second stage energy share search: operator profit gain over smart charging (solid, left axis) and the fixed credit $F^{\star}(\alpha)$ found in the first stage (dashed, right axis) versus the energy share $\alpha$ for the summer and winter rates (400 ports). Stars mark the best contracts.}
\label{fig:alpha}
\end{figure}

The energy share search is summarized in Table~\ref{tab:alpha} and Fig.~\ref{fig:alpha}. For each energy share, $F^{\star}(\alpha)$ is the smallest fixed credit that still gives all 400 owners at least their reserve. It falls as $\alpha$ rises, because the energy share covers more of the reserve. The profit gain is highest at $\alpha^{\star}=0.70$ with $F^{\star}=\$0.12$/EV-day. This is \$867 per week (21\,\%) more than a contract with only a fixed credit ($\alpha=0$). Above $\alpha^{\star}$, the fixed credit is already zero, so a larger share only gives more money to the owners and the gain falls. The best contract is therefore a real two part contract in which the energy share carries most of the owner payment. The winter curve in Fig.~\ref{fig:alpha} has the same shape and the same best share, $\alpha^{\star}=0.70$, but stays below zero, so no energy share makes V2G profitable in winter (Section~\ref{sec:seasonal}). All 400 owners take part. Their mean payoff is \$14.21 against a mean reserve of \$13.68, and 208 owners receive exactly their reserve. This is expected when the operator knows each owner's reserve, since the operator then keeps all the value above it.

With first come, first served port allocation, 4250 of the 5000 expected sessions are served, and no served vehicle has to wait. All 750 rejected sessions are V2G vehicles, while all commuter and fast charging demand is met. The 400 bidirectional ports are all in use at the busiest time, while the Level~2 and DCFC ports are only 18.9\,\% and 3.7\,\% used on average. The number of bidirectional ports, and not the total number of chargers, is therefore the limit. Section~\ref{sec:comparison} tests whether adding more bidirectional ports is worthwhile.

\subsection{Seasonal and Annual Assessment}
\label{sec:seasonal}

\begin{table}[t]
\centering
\caption{Profit Gain by Season and Annualized (400 Ports)}
\label{tab:seasonal}
\footnotesize
\setlength{\tabcolsep}{2.5pt}
\renewcommand{\arraystretch}{1.15}
\begin{threeparttable}
\begin{tabular}{@{}lrrr@{}}
\toprule
\textbf{Comparison} & \textbf{Summer (\$/wk)} & \textbf{Winter (\$/wk)} & \textbf{Annual (\$/yr)} \\
\midrule
Smart vs. uncontrolled  & 5658 & 4000 & 237478 \\
V2G vs. smart           & 4978 & $-$1095 & 48741 \\
\midrule
High peak cut, V2G (kW) & 4400 & 2177 & {--} \\
\bottomrule
\end{tabular}
\begin{tablenotes}[flush]\footnotesize
\item Annual values follow \eqref{eq:annual} with 17.4 summer and 34.7 winter weeks. The V2G high peak cut is measured against smart charging.
\end{tablenotes}
\end{threeparttable}
\end{table}

Table~\ref{tab:seasonal} repeats the comparison with the winter rates, in which the high peak demand rate falls from \$9.70 to \$4.30/kW-month and the low peak demand charge is removed. The value of V2G depends strongly on the season. In winter, V2G still lowers the high peak demand by 2177~kW, but the demand charge savings no longer cover the owner payments required by the participation constraint. Even the best winter contract ($\alpha^{\star}=0.70$, $F^{\star}=\$0.27$/EV-day) lowers the operator profit by \$1095 per week. Smart charging needs no owner payments and stays profitable in both seasons. Over a full year, smart charging alone gains \$237478, and V2G adds \$48741 on top of it if it also runs in winter. An operator would not run a program that loses money, so V2G is offered in summer only, which raises its annual gain to \$86761. Only this summer only value is used in the rest of the paper.

\subsection{Sensitivity Analysis}
\label{sec:sensitivity}

\begin{table}[t]
\centering
\caption{Sensitivity of the V2G Profit Gain Over Smart Charging (400 Ports)}
\label{tab:sensitivity}
\footnotesize
\setlength{\tabcolsep}{3.5pt}
\renewcommand{\arraystretch}{1.15}
\begin{threeparttable}
\begin{tabular}{@{}lcrrrr@{}}
\toprule
\textbf{Case} & $\boldsymbol{(\alpha^{\star},F^{\star})}$\tnote{a} & \textbf{Summer} & \textbf{Winter} & \textbf{Annual}\tnote{b} & \textbf{Prem.}\tnote{c} \\
 & & \textbf{\$/wk} & \textbf{\$/wk} & \textbf{\$/yr} & \textbf{\$/port} \\
\midrule
Base                     & (0.7, 0.12) & 4978 & $-$1095 & 86761  & 1808 \\
$c_{\mathrm{rep}}=70$    & (0.5, 0.09) & 7922 & 1695    & 196892 & 4102 \\
$c_{\mathrm{rep}}=130$   & (0.9, 0.31) & 1724 & $-$4272 & 30038  & 626 \\
Demand $\times0.75$      & (0.7, 0.12) & 1686 & $-$2220 & 29393  & 612 \\
Demand $\times1.25$      & (0.7, 0.12) & 8319 & 92      & 148184 & 3087 \\
No HP premium            & (0.7, 0.13) & 6038 & $-$1104 & 105231 & 2192 \\
$N_{\max}=2$             & (0.7, 0.07) & 4692 & $-$138  & 81778  & 1704 \\
$N_{\max}=4$             & (0.7, 0.21) & 4369 & $-$2617 & 76154  & 1587 \\
\bottomrule
\end{tabular}
\begin{tablenotes}[flush]\footnotesize
\item[a] Summer contract, $F^{\star}$ in \$/EV-day.
\item[b] V2G suspended in any season with a negative gain, weighted by \eqref{eq:annual}.
\item[c] Largest bidirectional premium per port, over a unidirectional port, that gives a 10 year simple payback with 2\,\% annual O\&M.
\end{tablenotes}
\end{threeparttable}
\end{table}

To test how much the results depend on the assumptions, the four inputs that drive the owner payments and the demand charge savings are changed one at a time. For each case, both seasons are solved again with the full energy share search. The battery replacement cost $c_{\mathrm{rep}}$, which scales the wear cost and the participation reserve, is changed from 98 to 70 and 130~\$/kWh. All demand charge rates, including the facilities charge, are scaled by 0.75 and 1.25. The 20\,\% high peak retail premium is removed, and the cycle limit $N_{\max}$ is set to 2 and 4. The results are shown in Table~\ref{tab:sensitivity}. The base case matches Sections~\ref{sec:results}--\ref{sec:seasonal}, and every owner meets the participation constraint in every case.

Two results hold in every case. First, V2G is profitable in summer, with weekly gains between \$1686 and \$8319. Second, the summer high peak reduction stays at 4400~kW, except with lower demand charges (4311~kW) and with the tighter cycle limit (3636~kW). Since 4400~kW equals 400 vehicles discharging at 11~kW each, the peak reduction is limited by the fleet's discharge power, not by its stored energy.

The winter loss in Section~\ref{sec:seasonal}, however, does not hold in every case. It turns into a gain of \$1695 per week when the replacement cost falls to 70~\$/kWh, because both the reserve and the wear payment fall with battery cost. It reaches break even when the demand charges rise by 25\,\%. The two most important inputs are therefore the battery replacement cost, which sets the cost of paying owners, and the demand charge level, which sets the value of lowering the peak. A 25\,\% change in demand charges shifts the summer gain by about \$3300 per week. This equals 25\,\% of the \$13290 per week of demand charge savings in the base case, so the gain changes linearly with the tariff. The retail premium has only a small effect, since it moves the smart charging and V2G cases by similar amounts. The cycle limit shows a seasonal trade off. Three cycles is close to the best choice in summer, since both two and four cycles lower the gain. In winter, two cycles cut the loss from \$1095 to \$138 per week, because the reserve owed to owners falls with the cycle allowance while the extra winter cycles bring little demand relief. A contract with a different cycle allowance in each season would therefore recover most of the winter loss.

Finally, smart charging also needs ports for the 400 long stay vehicles, so the cost that V2G must recover is only the extra price of a bidirectional port over a standard one. In the base case, V2G pays back in 10 years if this extra price is below \$1808 per port. Across the sensitivity cases, this limit ranges from \$612 to \$4102. Whether V2G is worthwhile at the campus therefore depends mainly on the bidirectional hardware price, the battery replacement cost, and the demand charge level. The contract itself matters less, because the second stage adjusts it to each case.

\subsection{Investment Comparison: V2G Versus Stationary BESS}
\label{sec:investment}

\begin{table}[t]
\centering
\caption{Investment Comparison Against Smart Charging (400 Ports)}
\label{tab:investment}
\footnotesize
\setlength{\tabcolsep}{2.5pt}
\renewcommand{\arraystretch}{1.15}
\begin{threeparttable}
\begin{tabular}{@{}lrrrr@{}}
\toprule
\textbf{Metric} & \textbf{V2G} & \textbf{V2G}\tnote{a} & \textbf{BESS} & \textbf{BESS}\tnote{b} \\
 & & \textbf{(summer)} & & \textbf{+ land} \\
\midrule
Equipment CAPEX (\$)          & 1800000 & 1800000 & 4709880 & 4709880 \\
Land (\$)                     & 0       & 0       & 0       & 591667 \\
Annual gain (\$/yr)           & 48741   & 86761   & 203278  & 203278 \\
Annual O\&M (\$/yr)           & 36000   & 36000   & 43220   & 43220 \\
Annual net (\$/yr)            & 12741   & 50761   & 160058  & 160058 \\
Simple payback (yr)           & 141.3   & 35.5    & 29.4    & 33.1 \\
Summer high peak cut (kW)     & 4400    & 4400    & 3095    & 3095 \\
CAPEX per kW cut (\$/kW)      & \textbf{409} & \textbf{409} & 1522 & 1713 \\
\bottomrule
\end{tabular}
\begin{tablenotes}[flush]\footnotesize
\item[a] V2G offered in summer only. V2G CAPEX conservatively assumes the full \$4500 per installed bidirectional EVSE and 2\,\% annual O\&M.
\item[b] 0.083~acre at the 120~MWh per acre density of an operating utility installation \cite{sdge2017escondido} and \$7.1\,M per acre for industrial land in the Los Angeles South Bay \cite{cornfield2023turnbridge}.
\end{tablenotes}
\end{threeparttable}
\end{table}
The stationary BESS described in Section~\ref{subsec:bess} is scheduled by the same LP, this time without V2G. In the summer week it raises the operator's profit by \$6732.93 compared with smart charging and lowers the high peak demand by 3095~kW, from 48580~kW to 45485~kW, at a battery wear cost of \$1715.53. Because the BESS has no owners to pay, it also stays profitable in winter, earning \$2475.44 per week. Over a full year it gains \$203278, more than V2G. A higher yearly gain, however, does not make the BESS a good investment for the campus, as the cost comparison in Table~\ref{tab:investment} shows.

First, the BESS is a much more expensive way to lower the peak. The 5~MW/10~MWh unit costs \$4.71\,M to buy, plus \$43220 per year to operate and maintain \cite{nrel2024atb}. The 400 bidirectional chargers cost \$1.80\,M. V2G lowers the summer high peak demand by 4400~kW at \$409 of investment per kW, while the BESS lowers it by only 3095~kW at \$1522 per kW, nearly four times as much.

Second, V2G uses parking spaces that already exist, but the BESS needs its own land next to the campus. The 120~MWh Escondido battery fits on about one acre, including its supporting equipment \cite{sdge2017escondido}. At the same density, a 10~MWh unit needs about 0.083~acres. Industrial land in the Los Angeles South Bay, the area around the airport, has sold for \$7.1\,M per acre \cite{cornfield2023turnbridge}, which adds \$0.59\,M, or 12.6\,\%, to the cost of the BESS. This is the lowest likely estimate. Recent planning guidelines allow 3~acres per 100~MWh once container spacing, power equipment, and safety clearances are included \cite{mercom2025ctuil}, which would raise the land cost to \$2.13\,M. In addition, the NFPA~855 fire code requires minimum distances between battery units and from nearby buildings, which are hard to provide at a crowded airport site \cite{nfpa855}. With land included, the time for the BESS to pay for itself grows from 29.4 years to between 33.1 and 42.7 years.

Third, the BESS never pays for itself within its lifetime in any of the cases studied. Large lithium-ion batteries are usually expected to last about 15 years \cite{nrel2024atb}, but the BESS needs 29.4 years to pay back even before land is counted. Across the demand charge and retail premium cases of Section~\ref{sec:sensitivity}, its yearly gain ranges from \$141514 to \$275183. Even in the best case, with demand charges 25\,\% higher, it needs 20.3 years to pay back without land.

Finally, the V2G numbers in Table~\ref{tab:investment} are cautious. They charge V2G the full cost of each charger, even though smart charging would need chargers for the same vehicles anyway. The cost V2G really adds is only the extra price of a bidirectional charger over a standard one. If V2G is charged only this extra price and is run in summer only, it pays back faster than the BESS with land whenever the extra price is below \$4320 per charger. With the higher land estimate, V2G pays back faster even at the full charger cost of \$4500 (35.5 years versus 42.7 years). As Section~\ref{sec:sensitivity} shows, an extra price below \$1808 per charger brings the V2G payback down to 10 years. The parked EVs therefore lower the peak more, for a fraction of the investment, without buying land or meeting battery siting rules. A stationary BESS is not a suitable investment for this campus under the current tariff.

\subsection{V2G Port Sizing: 400 Versus 800 Ports}
\label{sec:comparison}

\begin{table}[t]
\centering
\caption{V2G Port Sizing: 400 Versus 800 Bidirectional Ports (5000 Expected Sessions)}
\label{tab:comparison}
\footnotesize
\setlength{\tabcolsep}{6pt}
\renewcommand{\arraystretch}{1.15}
\begin{threeparttable}
\begin{tabular}{@{}l r r@{}}
\toprule
\textbf{Metric} & \textbf{400 ports} & \textbf{800 ports} \\
\midrule
Served sessions                      & 4,250 & 4,650 \\
V2G rejected                         & 750 & 350 \\
V2G utilization (\%)\tnote{a}        & 64.0 & 64.0 \\
V2G energy discharged (kWh/wk)       & 67,761 & 129,908 \\
\midrule
Summer contract $(\alpha^{\star},F^{\star})$ & (0.70, \$0.12) & (0.70, \$0.14) \\
Binding owners                       & 208/400 & 800/800 \\
Owner payments (\$/wk)               & 10,192 & 19,608 \\
High peak cut vs. smart (kW)         & 4,400 & 5,249 \\
\midrule
$\Delta\Pi$ vs. smart, summer (\$/wk)  & \textbf{4,978} & 4,091 \\
$\Delta\Pi$ vs. smart, winter (\$/wk)  & $-$1,095 & $-$4,951 \\
Annual net, summer only (\$/yr)\tnote{b} & \textbf{50,761} & $-$700 \\
\midrule
CAPEX (\$)                           & 1,800,000 & 3,600,000 \\
CAPEX per kW cut (\$/kW)             & \textbf{409} & 686 \\
Simple payback, summer only (yr)     & \textbf{35.5} & {--} \\
\bottomrule
\end{tabular}
\begin{tablenotes}[flush]\footnotesize
\item[a] Mean utilization; the peak is $N^{\mathrm{V2G}}/N^{\mathrm{V2G}}$ (fully committed) in both cases.
\item[b] V2G offered in summer only, after 2\,\% annual O\&M. CAPEX assumes \$4500 per installed bidirectional EVSE.
\end{tablenotes}
\end{threeparttable}
\end{table}

Table~\ref{tab:comparison} compares 400 and 800 bidirectional ports. The number of expected sessions is kept at 5000, so only the port count changes. The extra ports serve 400 more long stay vehicles and nearly double the discharged energy (from 67761 to 129908~kWh). The high peak reduction against smart charging grows from 4400~kW (9.1\,\%) to 5249~kW (10.8\,\%). However, the weekly profit gain over smart charging falls from \$4978 to \$4091 (17.8\,\% lower). There are two reasons. First, the demand charge savings show diminishing returns: once the first 400 ports have flattened the peak, the next 400 ports remove only 849~kW more. Second, every owner must be paid at least their reserve, so owner payments nearly double, from \$10192 to \$19608 per week (+92\,\%), and all 800 owners are paid exactly their reserve. The extra ports therefore cost more in owner payments than they save in demand charges. In winter, the weekly loss grows from \$1095 to \$4951. As a result, even with V2G run in summer only, the 800 port case does not cover its O\&M cost, and its capital cost per kW of peak reduction rises from \$409 to \$686. The 400 port case is therefore the better choice. In both cases, all V2G ports are in use at the busiest time, the mean V2G port use is 64.0\,\%, and every rejected session is a V2G vehicle (750 with 400 ports and 350 with 800 ports). The number of ports limits how many vehicles can join, but the number that is profitable is set by the demand charges and the owner payments, not by the port count.

\subsection{Representative Vehicle Dispatch and Compensation}

\begin{figure}[t]
\centering
\includegraphics[width=\columnwidth]{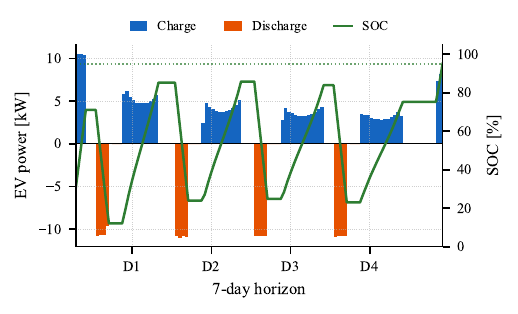}
\caption{Charge, discharge, and state of charge trajectory of a representative long stay V2G vehicle over the 7 day horizon.}
\label{fig:long_stay_ev}
\end{figure}

Fig.~\ref{fig:long_stay_ev} shows how the contract works for one long stay vehicle. The vehicle has a 75~kWh battery and stays parked for 111 hours (4.62 days). It arrives at 31.1\,\% state of charge and leaves at its 95.0\,\% target. Over the week, 241.7~kWh is charged and 172.6~kWh is sent back to the campus, which equals 2.88 full cycles, with peak rates of 10.6~kW charging and 11.0~kW discharging. The wear from this cycling is valued at \$11.72 and is fully paid back, so the owner loses nothing by taking part. On top of the wear payment, the owner receives \$14.11 from the energy share and \$0.56 from the fixed credit. The net payoff of \$14.67 equals the owner's reserve, so the participation constraint is tight for this vehicle: the owner is paid just enough to take part, and the operator keeps the rest. Fig.~\ref{fig:long_stay_ev} also shows that discharge happens in the daily high peak windows and is always followed by recharging, so the departure target is met while the demand charge savings are still delivered.

\section{Conclusion}
This paper modeled an airport campus where long stay EV parking, airport buildings, and an AI data center share one grid connection. A two stage method was proposed that schedules V2G and designs the payment to EV owners at the same time. The method makes sure that every owner is paid enough to take part, pays for battery wear, and respects the number of available chargers.

With 400 V2G chargers, V2G raised the operator's profit by \$4978 per week in summer compared with smart charging and lowered the high peak demand by 4400~kW (9\,\%). The best contract gives owners 70\,\% of the energy value plus a small daily payment, and earns 21\,\% more than a fixed payment alone. All 400 owners take part, and no served vehicle has to wait for a charger. Doubling the ports to 800 lowered the peak by a further 849~kW, but reduced the profit, since the owner payments grew faster than the demand charge savings.

The value of V2G depends on the season. In winter, the lower demand charges no longer cover the owner payments, so V2G is best offered in summer only, which gives an annual gain of \$86,761. The sensitivity analysis confirmed the summer benefit in every case and showed that V2G pays back within 10 years if a bidirectional charger costs less than about \$1800 more than a standard charger. A 5~MW/10~MWh stationary battery earns more per year, but costs nearly four times as much per kW of peak reduction, needs land that adds 13\,\% to 45\,\% to its cost, and does not pay back within its lifetime. V2G is therefore the better choice for lowering the summer peak at this campus. Future work will add other revenue sources, such as ancillary services.

\bibliographystyle{IEEEtran}
\bibliography{BibliographyFile}

@techreport{bnef2023battery,
   author      = {{BloombergNEF}},
   title       = {Electric Vehicle Outlook 2023},
   institution = {BloombergNEF},
   year        = {2023}
}

@article{orfanoudakis2024ev2gym,
  title={Ev2gym: A flexible v2g simulator for ev smart charging research and benchmarking},
  author={Orfanoudakis, Stavros and Diaz-Londono, Cesar and Y{\i}lmaz, Yunus Emre and Palensky, Peter and Vergara, Pedro P},
  journal={IEEE Transactions on Intelligent Transportation Systems},
  volume={26},
  number={2},
  pages={2410--2421},
  year={2024},
  publisher={IEEE}
}

@article{lee2023vehicle,
  title={Vehicle-to-grid optimization considering battery aging},
  author={Lee, Chih Feng and Bjurek, Kalle and Hagman, Victor and Li, Yang and Zou, Changfu},
  journal={IFAC-PapersOnLine},
  volume={56},
  number={2},
  pages={6624--6629},
  year={2023},
  publisher={Elsevier}
}

@online{Jason2025BatteryChemistry,
  author       = {Jason},
  title        = {How EV Battery Chemistry Affects Performance and Lifespan},
  year         = {2025},
  month        = feb,
  day          = {28},
  organization = {Midtronics},
  url          = {https://www.midtronics.com/blog/how-ev-battery-chemistry-affects-performance-lifespan/},
  note         = {Accessed: Jul. 17, 2026}
}

@misc{LADWP_EV_Rates,
  author       = {{Los Angeles Department of Water and Power (LADWP)}},
  title        = {Electric Vehicle Commercial Charging Service -- Monthly Rates},
  year         = {2026},
  url          = {https://www.ladwp.com/account/customer-service/electric-rates/special-commercial-industrial-rates/electric-vehicle-commercial-charging-service-monthly-rates},
  note         = {Accessed: Jul. 20, 2026}
}

@article{wang2025electric,
  title={Electric Vehicle Scheduling Model With Strategic Siting and Incentive Pricing Considering Participation Willingness},
  author={Wang, Yanjia and Xie, Da and Gu, Chenghong and Zhao, Pengfei and Wang, Xitian},
  journal={IEEE Transactions on Smart Grid},
  year={2025},
  publisher={IEEE}
}

@inproceedings{rahman2023making,
  title={Making a virtual power plant out of privately owned electric vehicles: From contract design to scheduling},
  author={Rahman, Saidur and Sales-Ortiz, Javier and Ardakanian, Omid},
  booktitle={Proceedings of the 14th ACM International Conference on Future Energy Systems},
  pages={459--472},
  year={2023}
}

@techreport{thingvad2023potential,
author      = {Andreas Barnekov Thingvad and Alaa Farhat Noureddine and Markus Hvid Monin and Varnith Anna and Lasse Stenh{\o}j Ingvardsen and Kjeld N{\o}rregaard and Lea Kornbeck Askholm},
title       = {Potential of Smart Charging and V2G},
institution = {Hybrid Greentech, Danish Technological Institute, and Copenhagen Airports},  
type        = {ALIGHT Project White Paper},
number      = {WP4.3},
year        = {2023},
month       = oct,
note        = {Version 1.0, ISBN 978-87-975041-0-9},
url         = {https://www.teknologisk.dk/_/media/89220_ALIGHT-WP4_3_white_paper_about_smart_charging_and_V2G.pdf}
}

@article{brinkel2020should,
  title={Should we reinforce the grid? Cost and emission optimization of electric vehicle charging under different transformer limits},
  author={Brinkel, NBG and Schram, WL and AlSkaif, TA and Lampropoulos, Ioannis and Van Sark, WGJHM},
  journal={Applied Energy},
  volume={276},
  pages={115285},
  year={2020},
  publisher={Elsevier}
}

@article{brinkel2026can,
  title={Can grid tariffs be fair and efficient? A comprehensive evaluation of tariff designs for smart electric vehicle integration},
  author={Brinkel, Nico and van Montfoort, Floris},
  journal={Energy Economics},
  pages={109407},
  year={2026},
  publisher={Elsevier}
}

@techreport{SFO2023EVLoadAnalysis,
  author       = {{San Francisco International Airport}},
  title        = {SFO Electric Vehicle Charging Load Analysis: Final Report},
  institution  = {San Francisco International Airport (SFO)},
  year         = {2023},
  month        = jan,
  type         = {Final Report},
  url          = {https://sustainability.flysfo.com/wp-content/uploads/2024/03/2023-SFO-Electric-Vehicle-Charging-Load-Analysis.pdf},
  note         = {Accessed: Jul. 24, 2026}
}

@inproceedings{al2024comparative,
  title={Comparative analysis of MILP and RL algorithms for EV charging scheduling at airport parking},
  author={Al Mehadi, Abdullah and Watson, Orion and Kamineni, Abhilash},
  booktitle={2024 IEEE Kansas Power and Energy Conference (KPEC)},
  pages={1--6},
  year={2024},
  organization={IEEE}
}

@article{vereno2025utilizing,
  title={Utilizing autonomous airport parking facilities as virtual power plants to provide frequency containment reserve},
  author={Vereno, Dominik and Sch{\"u}tz, Ren{\'e} and Polanec, Katharina and Maldonato, Francesco and Mielke-Sulz, Florian and K{\"o}gler, Damien and Nicodemus, Rolf and Weber, Johanna and Gross, Jounes-Alexander and Neureiter, Christian and others},
  journal={ACM SIGENERGY Energy Informatics Review},
  volume={4},
  number={4},
  pages={49--64},
  year={2025},
  publisher={ACM New York, NY, USA}
}

@article{guo2023infrastructure,
  title={Infrastructure planning for airport microgrid integrated with electric aircraft and parking lot electric vehicles},
  author={Guo, Zekun and Li, Bozheng and Taylor, Gareth and Zhang, Xin},
  journal={Etransportation},
  volume={17},
  pages={100257},
  year={2023},
  publisher={Elsevier}
}

@article{yan2026optimal,
  title={Optimal vehicle fleet and charging strategy for electric ground service equipment at airports based on flight schedule},
  author={Yan, Na and Tang, Tieqiao and He, Jia and Lin, Yu},
  journal={Transport Policy},
  pages={104123},
  year={2026},
  publisher={Elsevier}
}

@article{fan2026co,
  title={From Co-Location to Coordination: EV Aggregator Dispatch for Voltage Regulation of AI Data Center-Integrated Distribution Systems},
  author={Fan, Yaoyu and Tao, Xiaojie and Ye, Zhaoyi and Kumar, Krish and Gadh, Rajit},
  journal={Available at SSRN 7032060},
  year={2026}
}

@inproceedings{al2026benchmarking,
  title={Benchmarking MILP Guided BC-PPO for Grid Constrained EV Charging and V2G Charging},
  author={Al Mehadi, Abdullah and Sultana, Arifa and Morgan, Jackson and Harper, Mario and Wang, Hongjie and Zane, Regan},
  booktitle={2026 IEEE Green Technologies Conference (GreenTech)},
  pages={1--6},
  year={2026},
  organization={IEEE}
}

@inproceedings{al2025milp,
  title={An MILP-Based Optimization Framework for Integrated Charging and Ships-to-Grid Energy Management: A Case Study of Lake Travis, Texas},
  author={Al Mehadi, Abdullah and Morankar, Chinmay},
  booktitle={2025 IEEE Electric Ship Technologies Symposium (ESTS)},
  pages={330--335},
  year={2025},
  organization={IEEE}
}

@article{li2026collaborative,
  title={Collaborative optimization scheduling strategy of electric vehicle load cluster and data center for multi-station integration},
  author={Li, Siwei and Yuan, Shuai and Guo, Shulin and Yu, Jiancheng and Pei, Xiaolu and Zhang, Zhida},
  journal={IEEE Access},
  year={2026},
  publisher={IEEE}
}

@article{sen2026consent,
  title={CONSENT: A Negotiation Framework for Leveraging User Flexibility in Vehicle-to-Building Charging under Uncertainty},
  author={Sen, Rishav and Liu, Fangqi and Talusan, Jose Paolo and Pettet, Ava and Suzue, Yoshinori and Bailey, Mark and Mukhopadhyay, Ayan and Dubey, Abhishek},
  journal={arXiv preprint arXiv:2601.01581},
  year={2026}
}

@article{calearo2022grid,
  title={Grid integration of electric vehicles: battery degradation and user needs},
  author={Calearo, Lisa},
  year={2022},
  publisher={DTU Wind and Energy Systems}
}

@misc{sdge2017escondido,
  author       = {{San Diego Gas \& Electric Company and Southern California Gas Company}},
  title        = {Pipeline Safety \& Reliability Project ({A}.15-09-013): Responses to {CPUC} Data Request No. 06 Follow Up},
  howpublished = {California Public Utilities Commission},
  month        = dec,
  year         = {2017},
  note         = {Item 3-17, Response 8: Escondido ESS is 120~MWh on approximately 1 acre including balance of plant},
  url          = {https://ia.cpuc.ca.gov/environment/info/ene/sandiego/Documents/Data%20Request%20No.%206.pdf},
  urldate      = {2026-10-12}
}

@misc{cornfield2023turnbridge,
  author       = {Greg Cornfield},
  title        = {Turnbridge Equities Pays \$7.1{M} Per Acre for Industrial Site in {LA}'s South Bay},
  howpublished = {Commercial Observer},
  month        = may,
  year         = {2023},
  url          = {https://commercialobserver.com/2023/05/la-turnbridge-equities-7-1m-per-acre-industrial/},
  urldate      = {2026-10-12}
}

@misc{mercom2025ctuil,
  author       = {{Mercom India}},
  title        = {{CTUIL} Proposes 3 Acre/100 {MWh} as Land Requirement for {BESS} Projects},
  month        = may,
  year         = {2025},
  url          = {https://mercomindia.com/ctuil-proposes-3-acre-100-mwh-as-land-requirement-for-bess-projects},
  urldate      = {2026-10-12}
}

@standard{nfpa855,
  author       = {{National Fire Protection Association}},
  title        = {{NFPA} 855: Standard for the Installation of Stationary Energy Storage Systems},
  year         = {2023},
  address      = {Quincy, MA}
}

@misc{nrel2024atb,
  author       = {{National Renewable Energy Laboratory}},
  title        = {Annual Technology Baseline 2024: Utility-Scale Battery Storage},
  year         = {2024},
  url          = {https://atb.nlr.gov/electricity/2024/data},
  urldate      = {2026-10-12}
}
\end{document}